\documentstyle{elsart}
\begin{document}
\begin{frontmatter}

\title{Black-Scholes option pricing within It\^{o} and 
Stratonovich conventions}
\author[ub]{J. Perell\'o},
\author[ub,gaesco]{J. M. Porr\`a},
\author[ub]{M. Montero} and 
\author[ub]{J. Masoliver}
\address[ub]{Departament de F\'{\i}sica Fonamental, 
Universitat de Barcelona, Diagonal, 647, 08028-Barcelona, Spain}
\address[gaesco]{Gaesco Bolsa, SVB, S.A., Diagonal, 429, 08036-Barcelona,
Spain} 
\date{\today}
\maketitle

\begin{abstract}
Options are financial instruments designed to protect investors
from the stock market randomness. In 1973, Fisher Black, Myron Scholes and
Robert Merton proposed a very popular option pricing method using
stochastic differential equations within the It\^o interpretation. Herein,
we derive the Black-Scholes equation for the option price using the
Stratonovich calculus along with a comprehensive review, aimed to
physicists, of the classical option pricing method based on the 
It\^{o} calculus. We show, as can be expected, that the Black-Scholes
equation is independent of the interpretation chosen. We nonetheless 
point out the many subtleties underlying Black-Scholes option pricing
method.  
\end{abstract}
\end{frontmatter}

\section{Introduction}

An European option is a financial instrument giving to its owner the right
but not the obligation to buy (European call) or to sell
(European put) a share at a fixed future date, the maturing time $T$, and
at a certain price called exercise or striking price $x_C$. In fact, this
is the most simple of a large variety of contracts
that can be more sophisticated. One of those possible extensions is the
American option which gives the right to exercise the option at any time
until the maturing time. In a certain sense, options are a security for
the investor thus avoiding the unpredictable
consequences of operating with risky speculative stocks.       

The trading of options and their theoretical study have been known for
long, although they were relative obscure and unimportant financial
instruments until the early seventies. It was then when options
experimented an spectacular development. The Chicago Board Options
Exchange, created in 1973, is the first attempt to unify options in one
market and trade them on only a few stock shares. The market rapidly
became a tremendous success and led to a series of innovations in option 
trading~\cite{cox1}. 

The main purpose in option studies is to find a fair and presumably 
riskless price for these instruments. The first solution to the problem
was given by Bachelier in 1900~\cite{bachelier}, and several option prices
were proposed without being completely satisfactory~\cite{smith}. However,
in the early seventies it was finally developed a complete option
valuation based on equilibrium theoretical hypothesis for speculative
prices. The works of Fisher Black, Myron Scholes~\cite{black} and
Robert Merton~\cite{merton00} were the culmination of this
great effort, and left the doors open for extending the option pricing
theory in many ways. In addition, the method has been proved to be
very useful for investors and has helped to option markets to have the
importance that they have nowadays in finance~\cite{cox1,smith}.  
             
The option pricing method obtains the so-called Black-Scholes equation
which is a partial differential equation of the same kind as the diffusion
equation. In fact, it was this similarity that led Black and Scholes to
obtain their option price formula as the solution of the diffusion
equation with the initial and boundary conditions given by the option
contract terms. Incidentally, these physics studies applied to economy
have never been disrupted and there still is a growing effort of the
physics community to understand the dynamics of finance from approaches
similar to those that tackle complex systems in
physics~\cite{bou,mandelbrot,arthur,lux,bak}. 

The economic ideas behind the Black-Scholes option pricing theory 
translated to the stochastic methods concepts are as follows. 
First, the option price depends on the stock price and this is a random 
variable evolving with time. Second, the efficient market
hypothesis~\cite{fama}, {\it i.e.}, the market incorporates
instantaneously any information concerning future market evolution,
implies that the random term in the stochastic equation must be
delta-correlated. That is: speculative prices are driven by white
noise~\cite{bou,lo}. It is known that any white noise can be written as a
combination of the derivative of the Wiener process and white shot noise
~\cite{skorohod}. In this framework, the Black-Scholes option pricing
method was first based on the geometric Brownian
motion~\cite{black,merton00}, and it was lately extended to include white
shot noise~\cite{merton0,cox}.

As is well known, any stochastic differential equation (SDE) driven by a
state dependent white noise, such as the geometric Brownian motion, is
meaningless unless an interpretation of the multiplicative noise term is
given. Two interpretations have been presented: It\^o~\cite{ito} and
Stratonovich~\cite{strato}. To our knowledge, all derivations of the
Black-Scholes equation starting from a SDE are based on the It\^o
interpretation. A possible reason is that
mathematicians prefer this interpretation over the Stratonovich's one,
being the latter mostly preferred among physicists. Nonetheless, as we try
to point out here, It\^o framework is perhaps more convenient for finance
being this basically due to the peculiarities of trading (see Sect. 4). In
any case, as Van Kampen showed some time ago~\cite{vankampen} no physical
reason can be attached to the interpretation of the SDE modelling price
dynamics. However, the same physical process results in two different SDEs
depending on the interpretation chosen. In spite of having different
differential equations as starting point, we will show that the resulting
Black-Scholes equation is the same regardless the interpretation of the
multiplicative noise term, and this constitutes the main result of the
paper. In addition, the mathematical exercise that represents this
translation into the Stratonovich convention provides a useful review,
specially to physicists, of the option pricing theory and the
``path-breaking" Black-Scholes method. 

The paper is divided in 5 sections. After the Introduction, a 
summary of the differences between It\^o and Stratonovich calculus is
developed in Section~2. The following section is devoted to explain
the market model assumed in Black-Scholes option pricing method. 
Section~4 concentrates in the derivation of the Black-Scholes equation
using both It\^{o} and Stratonovich calculus. Conclusions are drawn in
Section~5, and some technical details are left to appendices.

\section{It\^o vs. Stratonovich}

It is not our intention to write a formal discussion on the differences
between It\^o and Stratonovich interpretations of stochastic differential
equations since there are many excellent books and reviews on 
the subject~\cite{skorohod,vankampen,katja,gardiner}. However, we
will summarize those elements in these interpretations that change the
treatment of the Black-Scholes option pricing method. In all our
discussion, we use a notation that it is widely used among physicists.

The interpretation question arises when dealing with a multiplicative 
stochastic differential equation, also called multiplicative Langevin
equation,
\begin{equation}
\dot{X}=f(X)+g(X)\xi(t),
\label{lang}
\end{equation}
where $f$ and $g$ are given functions, and $\xi(t)$ is Gaussian white
noise, that is, a Gaussian and stationary random process with zero mean
and delta correlated. Alternatively, Eq.~(\ref{lang}) can be written in
terms of the Wiener process $W(t)$ as
\begin{equation}
dX=f(X)dt+g(X)dW(t),
\label{lang2}
\end{equation}
where $dW(t)=\xi(t)dt$. When $g$ depends on $X$, Eqs.~(\ref{lang}) and
(\ref{lang2}) have no meaning, unless an interpretation of the
multiplicative term $g(X)\xi(t)$ is provided. These different
interpretations of the multiplicative term must be given because,
due to the extreme randomness of white noise, it is not clear what value
of $X$ should be used {\it even during an infinitesimal timestep $dt$}.
According to It\^o, that value of $X$ is the one
before the beginning of the timestep, {\it i.e.}, $X=X(t)$, whereas
Stratonovich uses the value of $X$ at the middle of the timestep:
$X=X(t+dt/2)=X(t)+dX(t)/2$. 

Before proceeding further with the consequences of the above discussion,
we will first give a precise meaning of the differential of random
processes driven by Gaussian white noise and its implications. Obviously, 
the differential of any random process $X(t)$ is defined by
\begin{equation}
dX(t)\equiv X(t+dt)-X(t).
\label{differential}
\end{equation}
On the other hand, the differential $dX(t)$ of any random process is equal
(in the mean square sense) to its mean value if its variance is, at least,
of order $dt^2$~\cite{skorohod}: $\left\langle[dX(t)-
\langle dX(t)\rangle]^2\right\rangle=O(dt^{2})$. We observe that from now
on all the results of this paper must be interpreted in the mean square
sense. The mean square limit relation can be used to show
that $|dW(t)|^2=dt$~\cite{gardiner}. We thus have
from Eq.~(\ref{lang2}) that
\begin{equation}
|dX|^2=|g(X)|^2dt+O(dt^{2}),
\label{dx2a}
\end{equation}
and we symbolically write
\begin{equation}
dX(t)=O\left(dt^{1/2}\right). 
\label{dx2} 
\end{equation}

Let us now turn our attention to the differential of the product of two
random processes since this differential adopts a different expression
depending on the interpretation (It\^o or Stratonovich) chosen. In
accordance to Eq.~(\ref{differential}), we define
\begin{equation}
d(XY)\equiv[(X+dX)(Y+dY)]-XY. 
\end{equation}
This expression can be rewritten in many different ways. 
One possibility is
\begin{equation} 
d(XY)=\left(X+\frac{dX}{2}\right) dY + \left(Y+\frac{dY}{2}\right) dX,
\label{sdiff}
\end{equation}
but it is also allowed to write the product as
\begin{equation}
d(XY)=XdY+YdX+dXdY.
\label{idiff}
\end{equation}
Therefore, we say that the differential of a product reads in the
Stratonovich interpretation when 
\begin{equation}
d(XY)\equiv X_S dY + Y_S dX,
\label{sproduct}
\end{equation}
where 
\begin{equation}
X_S(t)\equiv X(t+dt/2)=X(t)+dX(t)/2,
\label{xs}
\end{equation}
and similarly for $Y_S(t)$. Whereas we say that
the differential of a product follows the It\^o interpretation when 
\begin{equation}
d(XY)\equiv X_I dY + Y_I dX + dXdY,
\label{iproduct}
\end{equation}
where 
\begin{equation}
X_I(t)\equiv X(t),
\label{xi}
\end{equation}
and $Y_I(t)\equiv Y(t)$. Note that Eq.~(\ref{sproduct}) formally agrees
with the
rules of calculus while Eq.~(\ref{iproduct}) does not. Note also that
Eqs.~(\ref{sproduct}) and (\ref{iproduct}) can easily be generalized to
the product of two functions, $U(X)$ and $V(X)$, of the random process
$X=X(t)$. Thus
\begin{equation}
d(UV)=U(X_S)dV(X)+V(X_S)dU(X),
\label{sproducta}
\end{equation}
where $X_S$ is given by Eq.~(\ref{xs}), and $dV(X)=V(X+dX)-V(X)$ with an
analogous expression for $dU(X)$. Within It\^o convention we have
\begin{equation}
d(UV)=U(X)dV(X)+V(X)dU(X)+dU(X)dV(X).
\label{iproducta}
\end{equation}

Let us now go back to Eq.~(\ref{lang}) and see that one important
consequence of the above discussion is that the
expected value of the multiplicative term, $g(X)\xi(t)$, depends on
the interpretation given. In the It\^o interpretation, it is clear that
$\langle g(X)\xi(t)\rangle=0$ because the value of $X$ (and, hence the
value of $g(X)$) anticipates the jump in the noise. In other words, $g(X)$
is independent of $\xi(t)$. On the other hand, it can be
proved that within the Stratonovich framework the average of the
multiplicative term reads $g(X)g'(X)/2$ where the prime denotes the
derivative~\cite{gardiner}. The zero value of the average $\langle
g(X)\xi(t)\rangle$ makes It\^o convention very appealing because then the
deterministic equation for the mean value of $X$ only depends on the drift
term $f(X)$. In this sense, note that any multiplicative stochastic
differential equation has different expressions for the functions $f(X)$
and $g(X)$ depending on the interpretation chosen. In the Stratonovich
framework, a SDE of type Eq.~(\ref{lang2}) can be written as 
\begin{equation}
dX=f^{(S)}(X_S)dt+g^{(S)}(X_S)dW(t),
\label{lang2s}
\end{equation}
where $X_S=X+dX/2$. In the It\^o sense we have
\begin{equation}
dX=f^{(I)}(X_I)dt+g^{(I)}(X_I)dW(t),
\label{lang2i}
\end{equation}
where $X_I=X$. Note that $f^{(S)}$ and $f^{(I)}$ are not only
evaluated at different values of $X$ but are also different functions
depending on the interpretation given, and the same applies to $g^{(S)}$
and $g^{(I)}$. One can easily show from Eq.~(\ref{xs}) and
Eqs.~(\ref{lang2s})-(\ref{lang2i}) that, after keeping terms up to order
$dt$, the relation between $f_S$ and $f_I$ is~\cite{gardiner}
\begin{equation}
f^{(I)}(X)=f^{(S)}(X)-\frac 12 g^{(S)}(X)
\frac{\partial g^{(S)}(X)}{\partial X},
\label{itos}
\end{equation}
while the multiplicative functions $g^{(S)}$ and $g^{(I)}$ are equal
\begin{equation}
g^{(I)}(X)=g^{(S)}(X).
\label{itosa}
\end{equation}
Conversely, it is possible to pass from a Stratonovich SDE to an
equivalent It\^o SDE~\cite{gardiner}. Note that the difference between
both interpretation only affects the drift term given by the function $f$
while the function $g$ remains unaffected. In addition, we see
that for an additive SDE, {\it i.e.}, when $g$ is independent of $X$, the
interpretation question is irrelevant.

Finally, a crucial difference between It\^o and Stratonovich
interpretations appears when a change of variables is performed on the
original equation. Then it can be proved that, using Stratonovich
convention, the standard rules of calculus hold, but new rules appear when
the equation is understood in the It\^o sense. From the point of view of
this property, the Stratonovich criterion seems to be more convenient.
For the sake of completeness, we remind here what are the rules of change
of variables in each interpretation. Let $h(X,t)$ be an arbitrary function
of $X$ and $t$. In the It\^o sense, the differential of $h(X,t)$
reads~\cite{gardiner}
\begin{equation}
dh=\frac{\partial h(X,t)}{\partial X} dX+
\left[\frac{\partial h(X,t)}{\partial t}+
\frac 12g^2(X,t)\frac{\partial^2h(X,t)}{\partial X^2}\right]dt,
\label{dfi}
\end{equation}
whereas in the Stratonovich sense, we have the usual
expression~\cite{gardiner}
\begin{equation}
dh=\frac{\partial h(X_S,t)}{\partial X_S}dX+
\frac{\partial h(X_S,t)}{\partial t}dt,
\label{dfs}
\end{equation}
where
$$
\frac{\partial h(X_S,t)}{\partial X_S}=
\left.\frac{\partial h(X,t)}{\partial X}\right|_{X=X_S},
$$
and $X_S$ is given by Eq.~(\ref{xs}). 

Equation~(\ref{dfi}) is known as the It\^o's lema and it is extensively
used in mathematical finance 
books~\cite{lo,hull,baxter,wilmott,wilmott2,merton2}.

The information on the properties of the It\^o and Stratonovich
interpretation of SDE contained in this brief summary is sufficient to
follow the derivations of the next sections.

\section{Market model} 

Option pricing becomes a problem because market prices or indexes change
randomly. Therefore, any possible calculation of an option price is based
on a model for the stochastic evolution of the market prices. The first
analysis of price changes was given one hundred years ago by Bachelier 
who, studying the option pricing problem, proposed
a model assuming that price changes behave as an ordinary random
walk~\cite{bachelier}. Thus, in the continuum limit (continous time
finance~\cite{merton2}) speculative prices obey a Langevin equation. In
order to include the limited liability of the stock prices, {\it i.e.},
prices cannot be negative, Osborne proposed the geometric or
log-Brownian motion for describing the price changes~\cite{osborne}.
Mathematically, the market model assumed by Osborne can be written as a
stochastic equation of type Eq.~(\ref{lang2}):
\begin{equation}
  dR(t)=\mu dt+\sigma dW(t),
  \label{req}
\end{equation}
where $R(t)$ is the so-called return rate after a period $t$. Therefore,
$dR(t)$ is the infinitessimal relative change in the stock share
price $X(t)$ (see Eq.~(\ref{xteq}) below), $\mu$ is the average rate per
unit time, and $\sigma^2$ is the volatility per unit time of the rate
after a period $t$, {\it i.e.}, $\langle dR\rangle=\mu dt$ and
$\langle(dR-\langle dR\rangle)^2\rangle=\sigma^2dt$. There is no need to
specify an interpretation (It\^o's or Stratonovich's) for Eq.~(\ref{req})
because $\sigma$ is constant and we are thus dealing with an additive
equation. The rate is compounded continuously and, therefore, an initial
price $X_0$ becomes after a period  $t$:
\begin{equation}
X(t)=X_0 \exp[R(t)].
\label{xteq}
\end{equation}
This equation can be used as a change of variables to derive the SDE 
for $X(t)$ given that $R(t)$ evolves according to Eq.~(\ref{req}). 
However, as it becomes multiplicative, we have to attach the equation to
an interpretation. Indeed, using Stratonovich calculus (see
Eq.~(\ref{dfs})), it follows that $X(t)$ evolves according to the equation
\begin{equation}
dX=\mu X_Sdt+\sigma X_SdW(t),
\label{xeqst}
\end{equation}
where $X_S=X+dX/2$. In the It\^o sense (see Eq.~(\ref{dfi})), the equation
for $X(t)$ becomes
\begin{equation}
dX=\left(\mu+\sigma^2/2\right)Xdt+\sigma XdW(t).
\label{xeqit}
\end{equation}
Therefore, the Langevin equation for $X(t)$ is different depending on the
sense it is interpreted. The main objective of this paper
is to show that no matter which equation is used to derive the
Black-Scholes equation the final result turns out to be the same.

Before proceeding further, we point out that 
the average index price after a time $t$ is $\langle X(t)\rangle=X_0
\exp(\mu+\sigma^2/2)t$, regardless the convention being used. In
fact, the independence of the averages on the interpretation used holds
for moments of any order~\cite{vankampen,katja,gardiner}.

\section{The Black-Scholes equation}

There are several different approaches for deriving the Black-Scholes
equation starting from the stochastic differential equation point of 
view. These different derivations only differ in the way the portfolio
({\it i.e.}, a collection of different assets for diversifying away
financial risk) is defined~\cite{black,merton2,merton,harrison}. In order
to get the most general description of the concepts underlying in the
Black-Scholes theory, our portfolio is similar to the one proposed by
Merton~\cite{merton}, and it is based on one type of share whose
price is the random process $X(t)$. The portfolio is 
compounded by a certain amount of shares, $\Delta $,
a number of calls, $\Psi $, and, finally, a quantity of 
riskless securities (or bonds) $\Phi $. We also assume that short-selling,
or borrowing, is allowed. Specifically, we own a certain number of calls
worth $\Psi C$ dollars and we owe $\Delta X+\Phi B$ dollars. In this case,
the value $P$ of the porfolio reads
\begin{equation}
P=\Psi C-\Delta X-\Phi B,
\label{port}
\end{equation}
where $X$ is the share stock price,
$C$ is the call price to be determined, and $B$ is the
bond price whose evolution is not random and is described according to the
value of $r$, the risk-free interest rate ratio. That is 
\begin{equation}
dB=rBdt.
\label{bond}
\end{equation}
The so-called ``portfolio investor's strategy''~\cite{baxter} 
decides the quantity to be invested in every asset according to its stock
price at time $t$. This is the reason why the asset amounts 
$\Delta,\Psi,$ and $\Phi$ are functions of stock price and time,
although they are ``nonanticipating" functions of the stock price. This
somewhat obscure concept is explained in the Appendix A. All derivations
of Black-Scholes equation assume a
``frictionless market'', that is, there are no transaction costs for each
operation of buying and selling~\cite{black}. 

According to Merton~\cite{merton} we assume that, by short-sales, or
borrowing, the portfolio ~(\ref{port}) is constrained to require net zero
investment, that is, $P =0$ for any time~$t$~\cite{footnote}. Then,
from Eq.~(\ref{port}) we have
\begin{equation}
C=\delta_nX+\phi_nB,
\label{net}
\end{equation}
where, $\delta_n\equiv\Delta/\Psi$ and $\phi_n\equiv\Phi/\Psi$ are
respectively the number of shares per call and the number of bonds per
call. As we have mentioned above, $\delta_n$ and $\phi_n$ are
nonanticipating functions of the stock price (see Appendix A). Note that
Eq.~(\ref{net}) has an interesting economic meaning, since tells us that
having a call option is equivalent to possess a certain number, $\delta_n$
and $\phi_n$, of shares and bonds thus avoiding any arbitrage
opportunity~\cite{footnote}.
Equation~(\ref{net}), which is called ``the replicating
portfolio"~\cite{lo,baxter,wilmott}, is
the starting point of our derivation that we separate into two subsections
according to It\^o or Stratonovich interpretations. 

\subsection{The Black-Scholes equation derivation (It\^o)}

We need first to obtain, within the It\^o interpretation, the
differential of the call price $C$. This is done in the Appendix B and we
show there that 
\begin{equation}
dC=\delta dX+\phi dB+Xd\delta_n+Bd\phi_n+O(dt^{3/2}),
\label{IC}
\end{equation}
where the relationship between $\delta$, $\phi$ and $\delta_n$, $\phi_n$
is given in Appendix A ({\it cf.} Eq.~(\ref{a1})). We assume we follow a
``self-financing strategy"~\cite{harrison}, that is,
variations of wealth are only due to capital gains and not to the
withdrawal or infusion of new funds. In other words, we increase
[decrease] the number of shares by selling [buying] bonds in the same
proportion. We then have (see Appendix A for more details)
\begin{equation}
Xd\delta_n=-Bd\phi_n,
\label{sfinancing}
\end{equation}
and Eq.~(\ref{IC}) reads
\begin{equation}
dC=\delta dX+\phi dB.
\label{ICaa}
\end{equation}
Moreover, from Eqs.~(\ref{bond})-(\ref{net}) one can easily show that
$$
\phi dB=r(C-\delta X)dt+O(dt^{3/2}),
$$
({\it cf.} Eq.~(\ref{dx2}) and Eq.~(\ref{a1}) of Appendix A). Therefore,
\begin{equation}
dC=\delta dX+r(C-\delta X)dt+O(dt^{3/2}).
\label{ICa}
\end{equation}
On the other hand, since the call price $C$ is a function of share price
$X$ and time, $C=C(X,t)$, and $X$ obeys the (It\^o) SDE (\ref{xeqit}),
then $dC$ can be evaluated from the It\^o lemma (\ref{dfi}) with the
result
\begin{equation}
dC=\left(\frac{\partial C}{\partial t}+\frac 12
\sigma^2X^2\frac{\partial^2C}{\partial X^2}\right)dt+
\frac{\partial C}{\partial X}dX.
\label{ICb}
\end{equation}
Substituting Eq.~(\ref{ICa}) into Eq.~(\ref{ICb}) yields
\begin{equation}
\left(\delta-\frac{\partial C}{\partial X}\right)dX=
\left[\frac{\partial C}{\partial t}-r(C-\delta X)+
\frac 12\sigma^2X^2\frac{\partial^2C}{\partial X^2}\right]dt.
\label{ICc}
\end{equation}
Note that this is an stochastic equation because of its dependence on the
Wiener process enclosed in $dX$. We can thus turn Eq.~(\ref{ICc}) into a
deterministic equation that will give the call price functional
dependence on share price and time by equating to zero the term
multiplying $dX$. This, in turn, will determine the ``investor strategy",
that is the number of shares per call, the so called ``delta
hedging":
\begin{equation}
\delta=\frac{\partial C(x,t)}{\partial x}.
\label{dhedging}
\end{equation}
The substitution of Eq.~(\ref{dhedging}) into
Eq.~(\ref{ICc}) results in the Black-Scholes equation:
\begin{equation}
\frac{\partial C}{\partial t}= rC-rx\frac{\partial C}{\partial x}-
\frac 12(\sigma x)^2\frac{\partial^2C}{\partial x^2}.
\label{BS}
\end{equation}
A final observation, in Eqs.~(\ref{dhedging})-(\ref{BS}) we have set
$X=x$, since, as explained above, Eq.~(\ref{BS}) gives the functional
dependence of the call price $C$ on $X$ and $t$ regardless whether the
share price $X$ is random or not.

\subsection{The Black-Scholes equation derivation (Stratonovich)}

Let us now derive the Black-Scholes equation, assuming that the underlying
asset obeys the Stratonovich SDE (\ref{xeqst}). In the Appendix B we
present part of this derivation using the concept of nonanticipating
function within the Stratonovich interpretation. Nevertheless, here we
perform an alternative derivation that uses the It\^o interpretation as
starting point. We thus begin with Eq.~(\ref{ICa}) that we write in the
form
\begin{equation}
dC=\delta(X,t)dX(t)+r\left[C(X,t)-\delta(X,t)X\right]dt
+O(dt^{3/2}).
\label{SCa}
\end{equation}
Now, we have to express the function $\delta$ within
Stratonovich interpretation. Note that $X=X_S-dX/2$. Hence 
$\delta(X,t)=\delta(X_S-dX/2,t)$, whence
\begin{equation}
\delta(X,t)=\delta(X_S,t)-
\frac 12\frac{\partial\delta(X_S,t)}{\partial X_S} dX+O(dX^2).
\label{deltaS}
\end{equation}
Analogously $C(X,t)=C(X_S,t)+O(dX)$. Therefore, from
Eqs.~(\ref{SCa})-(\ref{deltaS}) and
taking into account Eq.~(\ref{dx2a}) we have
\begin{eqnarray}
dC=\delta(X_S,t)dX+\Biggl[rC(X_S,t)&-&rX_S\delta(X_S,t)
\nonumber\\
&-&\frac 12\sigma^2X_S^2\frac{\partial\delta(X_S,t)}
{\partial X_S}\Biggr]dt+O(dt^{3/2}).
\label{SCb}
\end{eqnarray}
On the other hand, $dC$ will also be given by Eq.~(\ref{dfs})
\begin{equation}
dC=\frac{\partial C(X_S,t)}{\partial t}dt+
\frac{\partial C(X_S,t)}{\partial X_S}dX,
\label{SCc}
\end{equation}
From these two equations we get
\begin{eqnarray}
\left[\delta(X_S,t)-\frac{\partial C(X_S,t)}{\partial X_S}\right]dX&=&
\Biggl[\frac{\partial C(X_S,t)}{\partial t}-rC(X_S,t)\nonumber\\
&+&rX_S\delta(X_S,t)
+\frac 12\sigma^2X_S^2\frac{\partial\delta(X_S,t)}
{\partial X_S}\Biggr]dt.
\label{SCd}
\end{eqnarray}
Again, this equation becomes non-stochastic if we set
\begin{equation}
\delta(X_S,t)=\frac{\partial C(X_S,t)}{\partial X_S}.
\label{dhedgings}
\end{equation}
In this case, the combination of Eqs.~(\ref{SCd})-(\ref{dhedgings}) agrees
with Eq.~(\ref{BS}). Although the call price is evaluated at a different
value of the share price, this is irrelevant for the reason explained
right after Eq.~(\ref{BS}). Therefore, the Stratonovich calculus results
in the same call price formula and equation than the It\^o calculus.

We have used the stochastic differential equation technique in order to
derive the option price equation. However, this is only one of the
possible routes. Another way, which was also proposed in the original
paper of Black and Scholes~\cite{black}, uses the Capital Asset Pricing
Model (CAPM)~\cite{sharpe} where, adducing equilibrium reasons in the
asset prices, it is assumed the equality of the so-called ``Sharpe ratio"
of the stock and the option respectively. The Sharpe ratio of an asset can
be defined as its normalized excess of return, therefore CAPM assumption
applied to option pricing reads~\cite{merton2}
$$
\frac{\alpha-r}{\sigma}=\frac{\alpha_C-r}{\sigma_C},
$$
where $\alpha=\langle dX/X\rangle$, $\sigma^2=\mbox{Var}(dX/X)$,
$\alpha_C=\langle dC/C\rangle$, and $\sigma_C^2=\mbox{Var}(dC/C)$. 
From this equality it is quite straightforward to derive the Black-Scholes
equation~\cite{black,merton2}. As remarked at the end of Sect. 3, moments
are independent of the interpretation chosen, we thus clearly see the
equivalence between It\^o and Stratonovich calculus for the Black-Scholes
equation derivation.

\subsection{The Black-Scholes formula for the European call}

For the sake of completeness, let us now finish the paper by shortly
deriving from Eq.~(\ref{BS}) the well-known Black-Scholes formula. Note
that the Black-Scholes equation is a backward parabolic differential
equation, we therefore need one ``final" condition and, in principle, 
two boundary conditions in order to solve it~\cite{carslaw}. In
fact, Black-Scholes equation is defined on the semi-infinite interval
$0\leq x<\infty$. In this case, since $C(x,t)$ is assumed to be
sufficiently well behaved for all $x$, we only need to specify one
boundary condition at $x=0$ (see \cite{wilmott2} and \cite{carslaw}),
although we specify below the boundary condition at $x=\infty$ as well.

We also note that all financial derivatives (options of any kind,
forwards, futures, swaps, etc...) have the same boundary conditions 
but different initial or final condition~\cite{wilmott}. Let us first
specify the boundary conditions. We see from the multiplicative character
of Eq.~(\ref{lang2}) that if at some time the price $X(t)$ drops to zero
then it stays there forever. In such a case, it is quite obvious that the
call option is worthless:
\begin{equation}
C(0,t)=0.
\label{BC1}
\end{equation}
On the other hand, as the share price increases without bound,
$X\rightarrow\infty$, the difference between share price and option price
vanishes, since option is more and more likely to be exercised and the
value of the option will agree with the share price, that is,
\begin{equation}
\lim_{x\rightarrow\infty}\frac{C(x,t)}{x}=1.
\label{BC2}
\end{equation}

In order to obtain the ``final" condition for Eq.~(\ref{BS}),
we need to specify the following two parameters: the expiration or
maturing time $T$, and the striking or exercise price $x_C$ that fixes the
price at which the call owner has the right to buy the share at time $T$.
If we want to avoid arbitrage opportunities, it is clear that the value of
the option $C$ of a share that at time $T$ is worth $x$ dollars must be
equal to the payoff for having the option~\cite{bachelier}. 
This payoff is either $0$ or the difference between share price at time
$T$ and option striking price, that is, $\max(x-x_C,0)$. Hence, the
``final" condition for the European call is
\begin{equation}
C(x,t=T)=\max(x-x_C,0).
\label{final}
\end{equation}

In the Appendix C we show that the solution to the problem given by
Eq.~(\ref{BS}) and Eqs.~(\ref{BC1})-(\ref{final}) is
\begin{equation}
C(x,t)=xN(d_1)-x_Ce^{-r(T-t)}N(d_2),
\label{formula}
\end{equation}
$(0\leq t\leq T)$, where 
$$
N(z)=\frac{1}{\sqrt{2\pi}}\int_{-\infty}^ze^{-u^2/2}du,
$$
is the probability integral,
$$
d_1=\frac{\ln(x/x_c)+(r+\sigma^2/2)(T-t)}{\sigma\sqrt{T-t}},
$$
and
$$
d_2=d_1-\sigma\sqrt{T-t}.
$$

\section{Conclusions}

We have updated the option pricing theory from the point of view of
a physicist. We have centered our analysis of option pricing to the
Black-Scholes equation and formula for the European call, extensions to
other kind of options can be straightforward in many cases and are found
in several good finance books~\cite{hull,baxter,wilmott,wilmott2,merton2}.
We have reviewed Black-Scholes theory using It\^o calculus, which is
standard to mathematical finance, with a special emphasis in explaining
and clarifying the many subtleties of the calculation. Nevertheless, we
have not limit ourselves only to review option pricing, but to derive, for
the first time to our knowledge, the Black-Scholes equation using the
Stratonovich calculus which is standard to physics, thus bridging the gap
between mathematical finance and physics. 

As we have proved, the Black-Scholes equation obtained using Stratonovich
calculus is the same as the one obtained by means of the It\^o calculus.
In fact, this is the result we expected in advance because It\^o and
Stratonovich conventions are just different rules of calculus. Moreover,
from a practical point of view, both interpretations differ only in the
drift term of the Langevin equation and the drift term does not appear in
the Black-Scholes equation and formula. But, again, we think that this
derivation is still interesting and useful for all the reasons explained
above.

\begin{ack}

This work has been supported in part by Direcci\'on General de
Investigaci\'on Cient\'{\i}fica y T\'ecnica under contract No. PB96-0188
and Project No. HB119-0104, and by Generalitat de Catalunya under contract
No. 1998 SGR-00015.

\end{ack}

\appendix

\section{Nonanticipating functions and self-financing strategy}

The functionals $\phi_n$ and $\delta_n$ representing normalized asset
quantities are nonaticipating functions with respect to the stock price
$X$. This means that these functionals are in some way independent of
$X(t)$ implying a sort of causality in the sense that unknown future
stock price cannot affect the present portfolio strategy.
The physical meaning of this translated to financial markets is: first buy
or sell according to the present stock price $X(t)$ and
right after the portfolio worth changes with variation of the prices
$dX$, $dB$, and $dC$. In other words, {\it the investor strategy does
not anticipate the stock price change}~\cite{smith,wilmott}.
Therefore, in the It\^o sense, the functionals $\delta_n$ and $\phi_n$
representing the number of assets in the portfolio solely depend on the
share price {\it right before} time $t$, {\it i.e.}, they do not depend on
$X(t)$ but on $X(t-dt)=X-dX$. That is,
\begin{equation}
\delta_n(X,t)\equiv\delta(X-dX,t),
\label{a1}
\end{equation}
and similarly for $\phi_n$ (recall that all equalities must be understood
in the mean square sense explained in Sect. 2). 

The expansion of Eq.~(\ref{a1}) yields (see Eq.~(\ref{dx2}))
$$
\delta_n(X,t)=\delta(X,t)-\frac{\partial\delta(X,t)}{\partial X}dX+O(dt),
$$
but from the It\^o lema (\ref{dfi}) we see that 
$$
\frac{\partial\delta(X,t)}{\partial X}dX=d\delta(X,t)+O(dt),
$$
and finally
\begin{equation}
\delta_n(X,t)=\delta(X,t)-d\delta(X,t)+O(dt).
\label{a2}
\end{equation}
Analogously,
\begin{equation}
\delta(X,t)=\delta_n(X,t)+d\delta_n(X,t)+O(dt),
\label{a3}
\end{equation}
and a similar expresion for $\phi(X,t)$. 

As to the self-financing strategy, Eq.~(\ref{sfinancing}), we observe that
$\delta(X,t+dt)$ is the number of shares we have at time $t+dt$, while 
$\delta(X-dX,t)$ is that number at time $t$. Therefore, 
$$
X(t)d\delta(X-dX,t)=[\delta(X,t+dt)-\delta(X-dX,t)]X(t)
$$
is the money we need or obtain for buying or from selling shares at time
$t$. Analogously, $B(t)d\phi(X-dX,t)$ is the money, needed or obtained at
time $t$, coming from bonds. If we follow a self-financing strategy, both
quantities are equal but with different sign, {\it i.e.},
\begin{equation}
X(t)d\delta(X-dX,t)=-B(t)d\phi(X-dX,t)
\label{a4}
\end{equation}
which agrees with Eq.~(\ref{sfinancing}).

\section{The differential of the option price}

Let us derive the differential of the call price, $dC$, using either It\^o
and Stratono\-vich interpretations. The starting point for both
derivations
is the replicating portfolio, Eq.~(\ref{net}),
\begin{equation}
C(X,t)=X(t)\delta_n(X,t)+B(t)\phi_n(X,t).
\label{ab5}
\end{equation}
Taking into account the It\^o product rule Eq.~(\ref{iproduct}), we have
\begin{eqnarray*}
dC=[\delta_n(X,t)+d\delta_n(X,t)]dX&+&[\phi_n(X,t)+d\phi_n(X,t)]dB
\\&+&X(t)d\delta_n(X,t)+B(t)d\phi_n(X,t),
\end{eqnarray*}
which, after using Eq.~(\ref{a3}), reads
\begin{eqnarray*}
dC=\delta(X,t)dX+\phi(X,t)dB
&+&X(t)d\delta_n(X,t)\\
&+&B(t)d\phi_n(X,t)+O(dt^{3/2}),
\end{eqnarray*}
and this agrees with Eq.~(\ref{IC}). 

Within the Stratonovich calculus, the differential of
Eq.~(\ref{ab5}) reads
\begin{equation}
dC=X_S(t)d\delta_n+B(t)d\phi_n+\delta_n(X_S,t)dX+\phi_n(X_S,t)dB.
\label{a5}
\end{equation}
From Eq.~(\ref{a1}) we have
\begin{equation}
\delta_n(X_S,t)=\delta(X_S,t)-\frac{\partial\delta(X_S,t)}
{\partial X_S}dX+O(dX^2),
\label{a6}
\end{equation}
and analogously for $\phi_n$. Substituting Eq.~(\ref{a6}) into
Eq.~(\ref{a5}), and taking into account
Eqs.~(\ref{dx2a})-(\ref{dx2}),~(\ref{xs}) and (\ref{bond}) we obtain
\begin{eqnarray*}
dC=[X(t)&+&dX/2]d\delta_n+B(t)d\phi_n+\delta(X_S,t)dX
\\&+&
\left[rB(t)\phi(X_S,t)-\sigma^2X_S^2\frac{\partial\delta(X_S,t)}
{\partial X_S}\right]dt+O(dt^{3/2}).
\end{eqnarray*}
But from Eq.~(\ref{a1}) and the self-financing strategy (\ref{a4}), we see
that $X(t)d\delta_n+B(t)d\phi_n=0$. Hence
\begin{eqnarray}
dC=\frac 12dXd\delta_n&+&\delta(X_S,t)dX\nonumber\\
&+&\left[rB(t)\phi(X_S,t)-\sigma^2X_S^2\frac{\partial\delta(X_S,t)}
{\partial X_S}\right]dt+O(dt^{3/2}).
\label{a7}
\end{eqnarray}
The substitution of the Stratonovich rule Eq.~(\ref{dfs}),
$$
d\delta_n=\frac{\partial\delta_n(X_S,t)}{\partial X_S}dX+
\frac{\partial\delta_n(X_S,t)}{\partial t}dt,
$$
yields
\begin{eqnarray}
dC=\delta(X_S,t)dX&+&
\left[rB(t)\phi(X_S,t)\right.\nonumber\\
&-&\left.\frac 12\sigma^2X_S^2\frac{\partial\delta(X_S,t)}
{\partial X_S}\right]dt+O(dt^{3/2}),
\label{a8}
\end{eqnarray}
where we have taken into account Eq.~(\ref{dx2a}) and the fact that
$\partial\delta_n/\partial X_S=\partial\delta/\partial X_S+O(dt^{1/2})$. 
Eq.~(\ref{a8}) agrees with Eq.~(\ref{SCb}) and the rest of the derivation
is identical to that of the main text.

\section{Solution to the Black-Scholes equation}

In this appendix we outline the solution to the Black-Scholes
equation~(\ref{BS}) under conditions~(\ref{BC1})-(\ref{final}). 

We first transform Eq.~(\ref{BS}) into a forward parabolic equation with
constant coefficients by means of the change of variables 
\begin{equation}
z=\ln(x/x_C),\qquad t'=T-t.
\label{b1}
\end{equation}
We have
\begin{equation}
\frac{\partial C}{\partial t'}=-r C(z,t')+
\left(r-\frac 12\sigma^2\right)\frac{\partial C}{\partial z}+
\frac 12\sigma^2\frac{\partial^2C}{\partial z^2},
\label{b2}
\end{equation}
$(-\infty<z<\infty,\, 0<t'<T)$. Moreover, the definition of a new
dependent variable:
\begin{equation}
u(z,t')=\exp\left[-\frac 12\left(1-\frac{2r}{\sigma^2}\right)z+
\frac 18\sigma^2\left(1+\frac{2r}{\sigma^2}\right)(T-t')\right]C(z,t'),
\label{b3}
\end{equation}
turns Eq.~(\ref{b2}) into the ordinary diffusion equation in an
infinite medium
\begin{equation}
\frac{\partial u}{\partial t'}=
\frac 12\sigma^2\frac{\partial^2u}{\partial z^2},
\label{b4}
\end{equation}
with a constant diffusion coefficient given by $\sigma^2/2$, and initial
condition:
\begin{eqnarray}
u(z,0)=x_C\exp\biggl[-\frac 12\biggl(1&-&\frac{2r}{\sigma^2}\biggr)z
\nonumber\\
&+&\frac 18\sigma^2\biggl(1+\frac{2r}{\sigma^2}\biggr)T\biggr]
\max\left(e^z-1,0\right).
\label{b5}
\end{eqnarray}
The solution of problem~(\ref{b4})-(\ref{b5}) is standard and reads
\cite{carslaw}
\begin{equation}
u(z,t')=\frac{1}{\sqrt{2\pi\sigma^2t'}}
\int_{-\infty}^{\infty}u(y,0)e^{-(z-y)^2/2\sigma^2t'}dy.
\label{b6}
\end{equation}
If we substitute the initial condition~(\ref{b5}) into the right hand side
of this equation and undo the changes of variables we finally obtain the
Black-Scholes formula Eq.~(\ref{formula}).

\end{document}